%% file: main.tex
\documentclass{article}
\usepackage{arxiv} 

\usepackage{graphicx}
\usepackage{subfigure}
\usepackage{arydshln}
\usepackage{amsmath}
\usepackage{amssymb}
\usepackage[round]{natbib}
\usepackage{url}

\newcommand{\figratio}{0.9}

\newcommand{\reffigure}[1]{Figure~\ref{#1}}

\newcommand{\cdt}[1]{{\small{#1}}}
\newcommand{\TRATE}{\cdt{T\_RATE}}
\newcommand{\SPREAD}{\cdt{SPREAD}}
\newcommand{\SCORE}{\cdt{SCORE}}
\newcommand{\NSCORE}{\cdt{NSCORE}}
\newcommand{\LL}{\cdt{LL}}
\newcommand{\NLL}{\cdt{NLL}}
\newcommand{\PRETEST}{\cdt{PRE-TEST}}
\newcommand{\POSTTEST}{\cdt{POST-TEST}}
\newcommand{\KEYBOARD}{\cdt{KEYBOARD}}
\newcommand{\TRACKER}{\cdt{TRACKER}}
\newcommand{\IKI}{\cdt{IKI}}
\newcommand{\PTT}{\cdt{PTT}}

\begin{document}

\title{Using Models of Baseline Gameplay to Design for Physical Rehabilitation}
\def\plainkeywords{Game; Rehabilitation; Interaction Design; User modelling.}
\renewcommand{\shorttitle}{Baseline Gameplay Models for Design}

\author{ 
Antoine Loriette\\
IRCAM, CNRS, Sorbonne Universit\'e\\
Paris, France\\
\texttt{antoine.loriette@gmail.com}\\
\And
Baptiste Caramiaux\\
Sorbonne Universit\'e, CNRS, ISIR\\
Paris, France\\
\texttt{caramiaux@isir.upmc.fr}\\
\And
Sebastian Stein\\
School of Computing Science, University of Glasgow\\
Glasgow, Scotland, United Kingdom\\
\texttt{sebastian.stein@glasgow.ac.uk}\\
\And
John H. Williamson\\
School of Computing Science, University of Glasgow\\
Glasgow, Scotland, United Kingdom\\
\texttt{johnh.williamson@glasgow.ac.uk}\\
}

\maketitle

\input{sections/abstract}
\input{sections/introduction}
\input{sections/background}
\input{sections/context}
\input{sections/experiment}
\input{sections/model}
\input{sections/discussion}

\bibliographystyle{abbrvnat}
\bibliography{main}

\end{document}

%% file: sections/abstract.tex

\begin{abstract}
Modified digital games manage to drive motivation in repetitive exercises needed for motor rehabilitation, however designing modifications that satisfy both rehabilitation and engagement goals is challenging. 
We present a method wherein a statistical model of baseline gameplay identifies design configurations that emulate behaviours compatible with unmodified play.
We illustrate this approach through a case study involving upper limb rehabilitation with a custom controller for a Pac-Man game. 
A participatory design workshop with occupational therapists defined two interaction parameters for gameplay and rehabilitation adjustments.
The parameters’ effect on the interaction was measured experimentally with 12 participants. 
We show that a low-latency model, using both user input behaviour and internal game state, identifies values for interaction parameters that reproduce baseline gameplay under degraded control. 
We discuss how this method can be applied to systematically balance gamification problems involving trade-offs between physical requirements and subjectively engaging experiences.
\end{abstract}

\keywords{\plainkeywords}

%% file: sections/introduction.tex

\section{Introduction}
The engaging nature of interactive systems, especially games, has the potential to improve rehabilitation sessions, as long as their interaction design aligns with patients' needs and capacities. One has to find and assess the relevant interaction design parameters that could be adapted to fit the rehabilitation requirements and specifications~\citep{Lopes2011}.

\textit{Gamification}~\citep{raczkowski2014making} seeks to incorporate ``\textit{video game elements in non-gaming systems to improve user experience and user engagement}''~\citep{Deterding2011}. The use of games has been shown beneficial for patients in a wide variety of rehabilitation contexts. Focusing on the upper limb, Sietsema et al.~\citep{Sietsema1993} showed that an interaction with a Simon game produced significantly more range of motion than a standard rote exercise for patients with traumatic brain injury.
However, when games are not initially intended for training purposes, the interaction between the game and the player must often be altered to meet rehabilitation goals. For example, Half-life 2~\citep{Halflife2} was partially controlled through a recumbent bicycle, by linking the pedalling speed to the move forward action of the playable character~\citep{Ketcheson2016}, for purposes of player exertion. 
Then additional parameterisation of the interaction is typically needed to accommodate specific patients ability~\citep{Barrett2016}. For instance, content from the game of Fruit-Ninja~\citep{FRUIT} was removed to cater for patients' reduced mobility during arm rehabilitation~\citep{Khademi2014}. These modifications are often introduced using heuristics~\citep{Pirovano2016, Munoz2018} or presets~\citep{Ketcheson2016, Chatta2015} but the extent to which they impact gameplay is difficult to predict.

In this paper, we propose to develop models of baseline gameplay, captured through game sessions from a control player group, to serve as a reference against which interaction modifications can be evaluated, alleviating the reliance on heuristics or presets.

Our contributions are twofold. First, we design a system for upper limb rehabilitation based on minimal game alterations and parametric interaction design. Second we validate a novel metric for predicting game difficulty using the designed system, based on models of baseline gameplay. 
These contributions benefit to the field of game design for rehabilitation by providing a new method for balancing game alterations compatible with a computational approach. 

This research is motivated by field work with occupational therapists leading to the development of a system for movement-based rehabilitation using the game of Pac-Man. A series of workshop and testing sessions grounded our design choices: interaction adaptation is performed with the modifications to the game's time rate and the optimisation of a newly designed input device physical properties. 
In addition, our contributions benefit to the field of Computational Interaction in HCI~\citep{Oulasvirta}, which relies on models to gain insight into interactive systems. We propose a model of baseline gameplay, constructed from low-level variables, that correlates with in-game score. We show that common measures of performance, such as in-game score, are prone to high latency and inter-user variability, and propose instead learning a statistical model of normative control behaviour.

The paper is structured as follows. In the next section we present previous work in modifying games for rehabilitation and related research to control input adaptation and user performance modelling. We then report the work carried out from an participatory workshops with occupational therapists and the resulting designed systems.
An experimental study involving 12 participants follows wherein we investigate how a new input modality affects the user experience and users' motions. Finally, to develop of a low-latency proxy for interaction difficulty, we detail a statistical modelling of collected data which is shown to correlate with game score.

%% file: sections/background.tex

\section{Related Work}

In line with our research objectives, we first review previous work in games for rehabilitation, and more precisely the modifications applied to the games to adapt to the rehabilitation specifications. Then we review related work on input gesture modelling in the context of gesture-based interaction. 

\subsection{Modifying Games for Rehabilitation}

A large body of research has been dedicated to developing \textit{Serious Games}~\citep{Michael2006,Susi2007} and design guidelines adapted to different patient's needs have been established for their creation, such as those targeting stroke patients and upper arm rehabilitation~\citep{Burke2009, Barrett2016}. More recently, repurposing commercial games has been proposed~\citep{Walther-Franks2013b} with the intent of leveraging their established gameplay. The distinction between serious games and commercial games can sometimes be ambiguous. Commercial games intentionally designed for physical exercise, such as those present on the Wii Fit platform, or serious games based on well-known commercial games~\citep{Burke2009, Khademi2014} are conceptually very similar. The gameplay from commercial games is however unparalleled and Barret et al.~\citep{Barrett2016} found in their review that most appreciated serious games were those closely inspired by commercial ones. 
While likely more engaging than serious games, commercial games are usually less flexible in their potential for adaptation. 

One commonality in the literature is the design of a new input control to adapt to different motor rehabilitation needs~\citep{Plaisant2000, Ketcheson2016, Khademi2014, Monedero2014, Chatta2015, Mandryk2013}. Traditional hand-held controllers have been replaced by or augmented with other systems. These sense parts of the user's body other than the hand, to solicit the motions required for physical rehabilitation while acting as input devices for control. Many studies have relied on the flexibility of optical technologies such as Microsoft Kinect~\citep{Bao2013,Chatta2015,Walther-Franks2013a}, Leap Motion controller~\citep{Khademi2014} or motions capture systems~\citep{Munoz2018}. Accelerometer based sensors from the Wii system were also employed~\citep{Walther-Franks2013a,Monedero2014} while instrumented versions of exercise equipment, such as recumbent bicycles~\citep{Ketcheson2016}, have also been used as input devices.
Likewise, our work relies on a custom gamepad to recruit arm gestures in an interaction with a commercial game originally designed for finger-based control.

Then, the actual gameplay can be changed to fit the rehabilitation requirements. Walther-Franks et al.~\citep{Walther-Franks2013b} proposed applying graphical overlays to unmodified games to fed back how well users performed rehabilitation-effective motions. Several systems have elaborated on this idea. 
Ketcheson et al.~\citep{Ketcheson2016} displayed the heart rate of the player and showed good results to support anti-sedentary levels of exertion, while Chatta et al.~\citep{Chatta2015} measured increased level of physical activity by employing graphical overlays to motivate players engaged with a commercial racing game. However, Chatta et al. also point out that this decouples exercise and in-game routine and, while it cleverly rely on ``inconvenient interactions''~\citep{Rekimoto2014}, such methods can go against common recommendations for serious game design~\citep{Sinclair2009}.
These are examples of fixed modifications.

Strategies of automatic adaptation to the players' skill level and difficulties have also be investigated, notably in the field of serious gaming  \citep{Hendrix2019, Burke2009, Sinclair2009}. These works highlight several challenges. 
Hendrix et al.~\citep{Hendrix2019} rightfully point out that games should not rely on presets, which require by nature a subjective player skill assessment, but adapt to an ideal challenge, Game knwoledge can be useful for that matter~\citep{Sinclair2009}. For instance, Burke et al.~\citep{Burke2009} mention that game difficulty can be tuned with the pace of the game, with a difficulty increasing along with the speed at which game elements move in the game.
A common approach to tune these parameters is to use heuristic rules. Mu{\~n}oz et al.~\citep{Munoz2018} used heuristic rules to adapt both game difficulty and rehabilitation goals for a serious game they designed, inspired by Pong, targeting an optimal pulse rate.
A set of rules linked to rehabilitation goals was also used by Pirovano et al.~\citep{Pirovano2016} to adapt the difficulty of custom games to specific users and a numerical objective was set to maintain an $80\%$ success rate in the game. 
Similar game design elements are used in our system, and the method we propose is aiming at removing the needs for using presets to adapt for game difficulty. Instead the baseline gameplay model sets the default challenge.

\subsection{Input control performance measure and modelling}

Modifying a control input or swapping one for another purposefully designed is likely to have an impact on performance.
A common measure of performance in games is the score, for example linked to the number of hits in First Person Shooter (FPS) games~\citep{Gerling2011}. Isokoski and Martin~\citep{Isokoski2007} compared a keyboard with a standard mouse, a wheel mouse, a trackmouse and a Xbox360 gamepad. The gamepad was outperformed by all other devices in the task of FPS target acquisition.
Another approach is to compute the information transmitted by the control inputs. Compared to a standard mouse, the Wii remote has been shown to perform poorly in terms of throughput, speed and error rate~\citep{Natapov2009}, while an Xbox gamepad performed equally well in tracking a target’s velocity~\citep{Klochek2006}. These differences can be explained by the differences in the limbs and muscles they recruit and is the topic of seminal research works. Card et al.~\citep{Card1991} compiled several Fitts' law experiments involving different limbs and showed a difference in available information throughput between neck, arm, wrist and finger -- with four-fold decrease between finger and arm from $38bit/s$ to $9.5bit/s$. Targeting more complex motions than pointing, Oulasvirta et al.~\citep{Oulasvirta2013} measured the information throughput in full body motions to provide researchers with a quantitative measure of control input capacity. Finally, to help designer in understanding the impact of new input methods on the user body, Bachynskyi et al.~\citep{Bachynskyi2015a} used a clustering approach to summarise results about 3d pointing. In summary, newly designed input controls are likely sub-optimal when compared to reference devices and methods to predict performance beyond pointing, in particular~\citep{Oulasvirta2013, Bachynskyi2015a}, exist but are heavy to deploy in workshop settings.

Models for gameplay interaction have also been developed. For example, Smith et al.~\citep{Smith2016} defined it as the player's raw input in contrast to methods that were only considering in-game states or event logs. In their work, they showed that models of gameplay can reliably identify player's identity based on their playstyle measured as ``input words'', i.e. strings of actions performed by the users. Patterns linking user identity to input behaviour was also found in the study from Dhakal et al.~\citep{Dhakal2018} in which they observed 136 millions keystroke of text input. In their analysis, several clusters identified users' typing behaviours. One of their performance measure included interkey interval, which is also a feature we rely on in this work.

Models of players have been used to replace humans in games by emulating their behaviour~\citep{Pfau2020}. They have also been used for game design generation to optimise game worlds with specific outcome~\citep{Karavolos2018}. When one has a direct access to player satisfaction, methods such as those proposed by Yannakakis et al.~\citep{Yannakakis2009} can be employed, in which they trained a model to control the parameters of a custom game in real-time using recorded player satisfaction in pairwise A/B tests and sampled possible game designs through genetic optimisation.
In contrast, the present work employs gameplay modelling for the purpose of identifying input configurations compatible with a baseline model.

\subsection{Summary}

The present work is, to the best of our knowledge, the first to develop baseline gameplay models of commercial games for adaptation to physical rehabilitation. In doing so, we tackle two challenges. The first challenge is finding parameters for the interaction loop that allows for rehabilitation and gameplay adaptation; this is addressed through co-design with occupational therapists. The second challenge is building a model of baseline gameplay from recorded game sessions; we solve this by proposing a statistical model, based on data captured in an experimental study, with four parameters that leverages player typing behaviour and recorded effects on game state.

%% file: sections/context.tex

\section{Motivation from a rehabilitation context}\label{sct:motivation}

The present work is grounded by field work with a team of occupational therapists (OTs) from the spinal injury unit of Queen Elizabeth University Hospital~\citep{Loriette2019}. 
They were interested in improving their process for upper limb reach rehabilitation, which has been less explored than lower limb treatment~\citep{Barrett2016}.

\subsection{Observations and Brainstorming workshops}

The present collaboration started with a discussion related to the nature of OTs' patients and moved onto focus on their rehabilitation practises.

Their target group gathers patients with high level of spinal cord injuries (SCI), typically with spinal lesions at levels C5 and C6~\citep{Jr1997}, for whom the arm functions are severely compromised~\citep{Anderson2004}. The induced partial or complete paralysis of specific arm muscles reduces patients' ability to perform coordinated motions.
During the first stages of rehabilitation in the weeks following injury, when fine motor control of the fingers is absent, functional rehabilitation goals were limited to whole arm motions over tabletops. 
In routine rehabilitation, OTs used two physical artefacts. A skateboard, onto which one arm of the patient is strapped, was used to elicit linear back and forth motions while supporting the patient's arm, and facilitating movement through the reduced friction of the skateboard's wheels.
The other artefact was a plastic tube rolled on its long dimension with patient's arm placed on top. A Velcro tape attached to the tube and to the table was sometimes used to increase friction as a mean to adapt task difficulty. For both devices, typical instructions from the OTs were for the patients to ``move towards the window'' or to ``reach forward'', involving four gestures in total: forward, backward, left, right on a tabletop. However, the simplistic, repetitive and artificial nature of these exercises led to low patient motivation, and they were described as ``dull, tedious and boring'' by the OTs.

Given the positive results reported in the literature regarding the use of games for rehabilitation~\citep{Barrett2016}, this involvement of games was discussed between us (HCI practitioners) and the OTs during a series of meetings. These discussions were helpful in two ways. They helped in defining the main requirements by the OTs for a system which elicits the same types of movements as the ones exerted using the two artefacts introduced previously. And they helped in understanding the extent to which they can be parameterised in order to calibrate the level of difficulty. A two-hours participatory workshop was then organised at the hospital with five OTs and two of the authors. The workshop was split equally into two parts. 

We first engaged in a brainstorming session to find mappings between game actions and rehabilitation movements. The list of rehabilitation movements was matched with a list of candidates for the games.
When prompted for simple games they were familiar with, the participants mentioned Whack-a-Mole, Bubble Witch, Pong and Frogger to name a few, all of which were arcade games.
The controls needed to implement their gameplay were then discussed. While some required four ways directional actions (e.g. Pac-Man), for aiming at a target for example, others only required button presses for triggering specific actions (e.g. Puzzle Bubble). 
To simplify the workshop complexity, it was decided that games with simple 4-ways controls should be focused on.

After the brainstorming, the second discussion addressed the parameterisation of specific exercises the OTs would like to see implemented by a gameplay. The movements that patients were producing when interacting with the skateboard, mentioned earlier, were used as a reference.
A paper prototype of actionable controls was created for the interaction to be played out on the desk around which the workshop took place. 
The instructions from the OTs were to perform long reaching motions with the arm in contact with a planar surface. The goal they pursued was for their patients to reach always further. The flexibility of the paper prototype highlighted that the possibility of actionable controls whose locations on the surface could be easily modified was important for exercise adaptation.

The outcome of these workshops was then threefold:
\begin{itemize}
    \item We identified that the game itself was not the focus for either parties (OTs and ourselves), as long as it is enjoyable for the patient and complies to the specifications given by OTs;
    \item We sketched/agreed on a control input with 4-ways actionable buttons to interface with various different games;
    \item We chose a rehabilitation parameter: the position of the gamepad's buttons.
\end{itemize}

\subsection{System description}

\begin{figure}[!ht]
\centering
\subfigure{\includegraphics[width=0.48\linewidth]{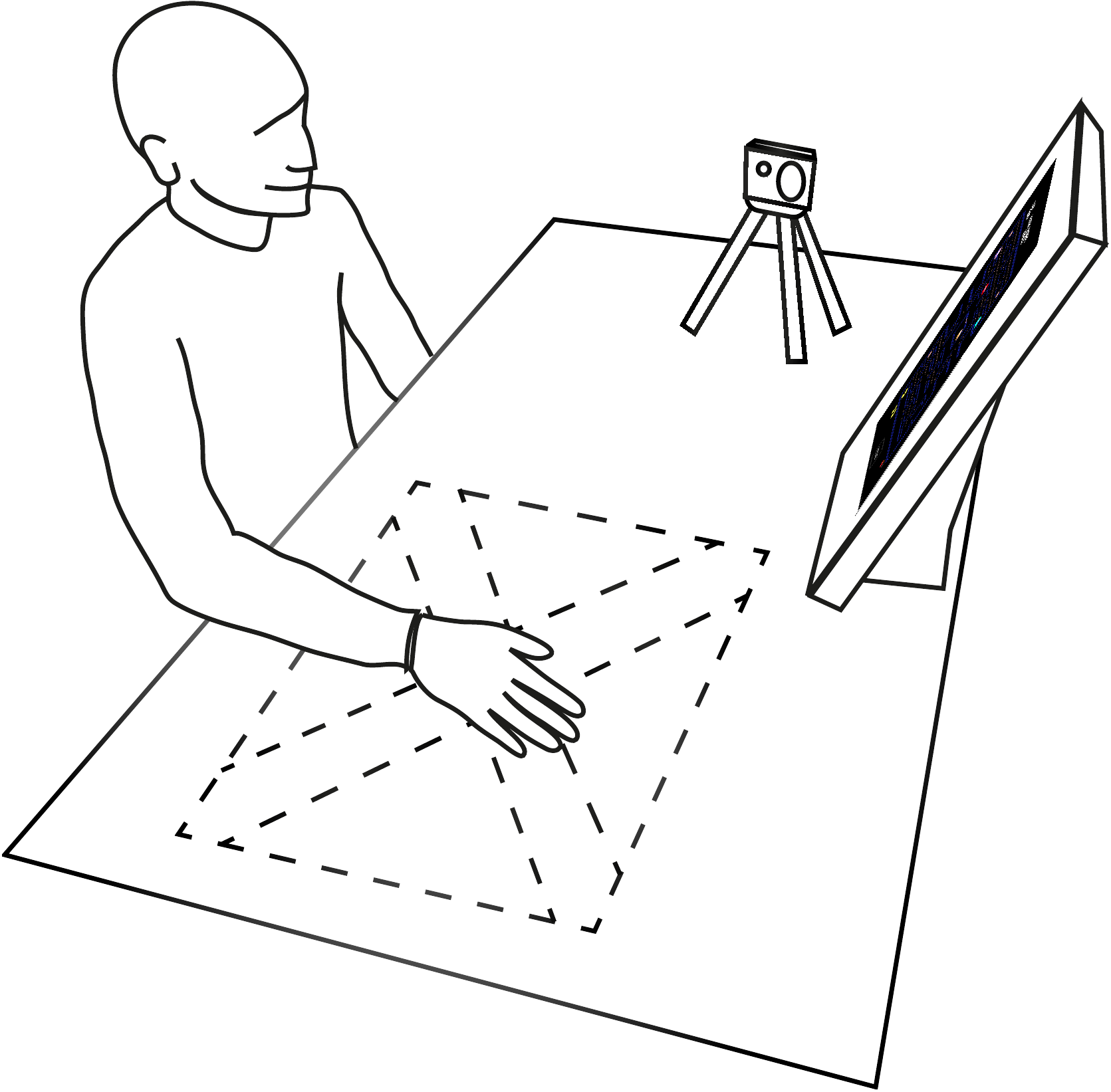}}
\subfigure{\includegraphics[width=0.32\linewidth]{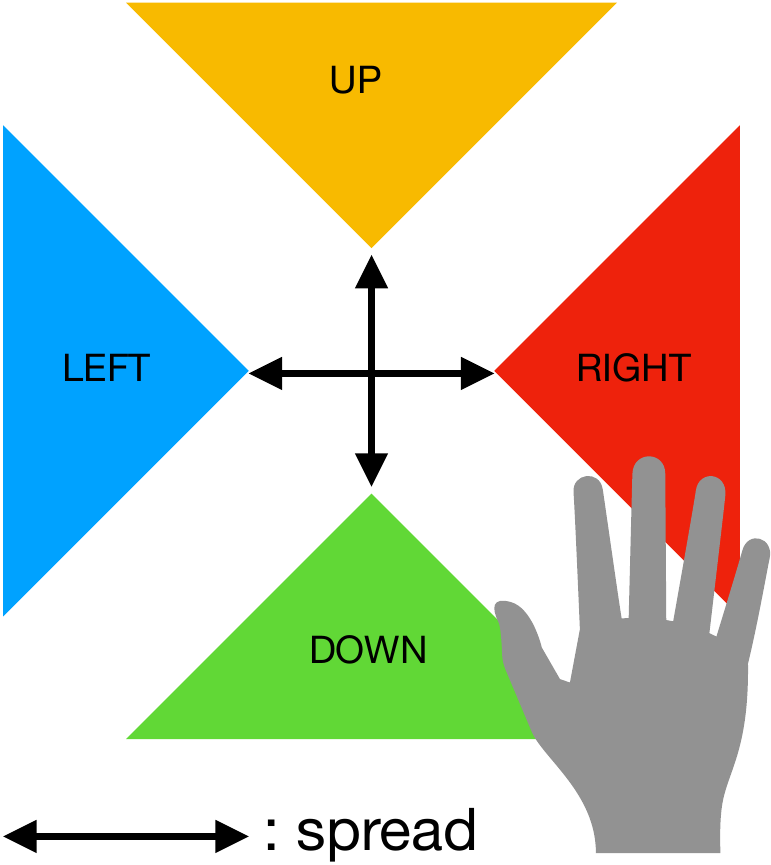}}
\caption{(left) Sketch of the proposed interaction. A user controls a digital game through four regions. Optical tracking tracks user's hand and detects when it enters gamepad regions. (right) The virtual gamepad, parameterised with the \textit{spread} variable, maps hand position to the commands UP, RIGHT, DOWN, LEFT.}
~\label{fig:proposed_interaction}
\end{figure}

The preliminary field work presented above has led to the prototype depicted in \reffigure{fig:proposed_interaction} (left). 
A patient is seated by a tabletop on top of which reaching motions are performed with one arm. An optical device both tracks the patient's hand and creates a virtual surface with four regions arranged in a cross shape. Each virtual button is activated when the hand of the user enters its respective region. This new input control is interfaced with a digital game running in front of the patient. The gestures required by the gameplay are meant to be similar to the one's needed for the patient's regular exercising.

As for the game, we chose Pac-Man~\citep{PACMAN} as test game. It is compatible with the four-way control and it is one of the most iconic arcade games where the rules are known by many. Regions of the virtual surface were mapped to directional commands (UP, RIGHT, DOWN, LEFT). Pac-Man always move forward at a given speed, the commands are only used to make directional changes.

The prototype was initially set up in the authors' lab on a 78cm wide office desk. We used an Optitrack system for the tracking of the participants hand. A single Infrared (IR) reflective marker, placed on the user's hand between the index and middle fingers, indicated the $(x,y)$ location of the hand on the surface. A dedicated computer ran the Optitrack system whose output was transmitted to another machine running the game in a browser. We used an opensource version of Pac-Man\footnote{See \url{https://github.com/masonicGIT/pacman} for the source code and a discussion about the accuracy of this remake.}. Translating $(x,y)$ coordinates of the hand to game commands was done with custom software on the machine running the game. The origin of the controls was indicated by a protruding tactile marker. Directional commands were triggered when the tracked hand location entered the triangular control regions, which were not visible to the user. Audio feedback indicated when commands were triggered with a \textit{bip}. Users would be equipped with headphones during use to ensure they could clearly hear this feedback and the in-game audio.

\subsection{Interaction parameterisation}

Based on the field work and the implemented system, interaction parameters have been identified in order to act upon the game difficulty and rehabilitation effectiveness.

A first parameter is the \textit{spread} parameter, denoted $r$, defined as the distance between the inner tips of the triangular actionable regions (\reffigure{fig:proposed_interaction} (right)). This parameter serves the purpose of varying task difficulty. This parameter was actually required by the OTs as a lever to adapt the game difficulty to the patient's needs. 

We conducted a first round of informal tests with members of our lab, revealing that the overall interaction difficulty was greatly increased due to the new input control modality (from button presses to reaching arm movements). We observed that the change in game difficulty incurred by the change of input control modality cannot be accommodated by the spread parameter alone. The latency and responsiveness of control is impacted by optical tracking delay and the effective reaching time necessary to move between regions. 

To counterbalance this effect, we identified a second interaction parameter controlling the speed of the game, called \textit{Time Rate}. 
This is similar to what games like Tetris use to set difficulty without altering other part of the gameplay mechanics and was for example recommented by Burke et al.~\citep{Burke2009} in their guidelines. It also has the benefit to limit game alterations to a unique and controllable parameter time rate $T_r$. By reducing the time rate, the number of actions per unit of time required for playing are reduced, facilitating the interaction. 

\begin{figure*}[!ht]
\centering
\includegraphics[width=\figratio\linewidth]{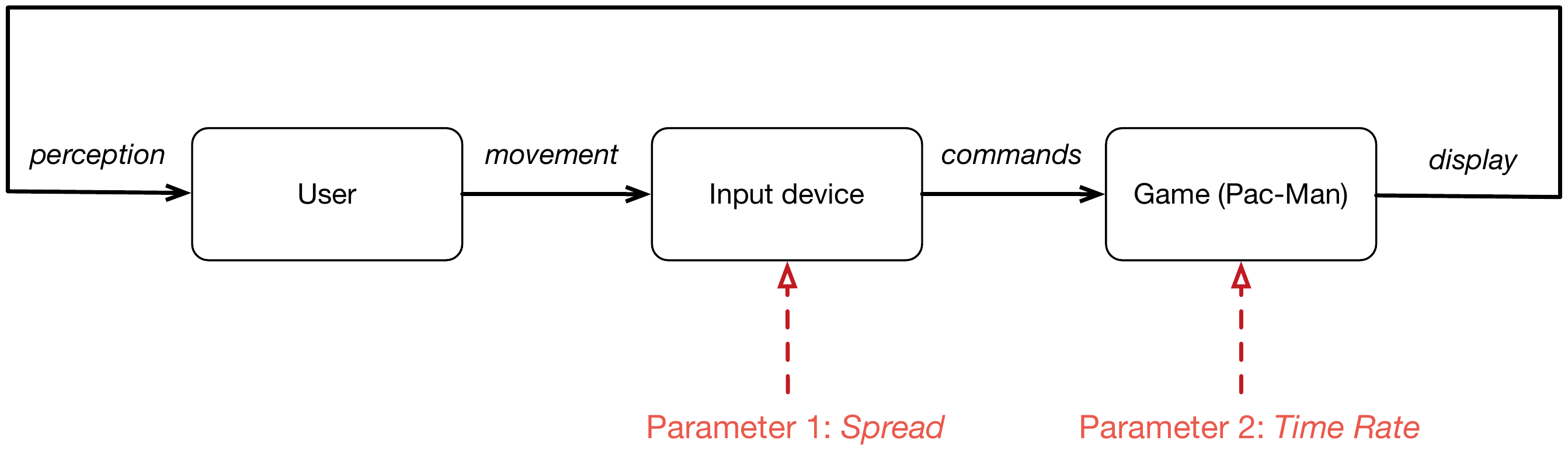}
\caption{Conceptual block diagram for the interaction showing the primary control loop. A user interact with a game through an input device. Design parameters spread and frame rate are used accommodate for rehabilitation goals and gameplay engagement.}
\label{fig:generalised_model}
\end{figure*}

The proposed interaction parameterisation can be represented as a block diagram (\reffigure{fig:generalised_model}). The perception loop includes the user, an input device and a game and produces engagement, depicted as a thick black line on the diagram. The interaction can be affected by two design parameters: the scale $r$, governing the distance that the participant has to travel with the arm in order to trigger a command; the time rate $T_r$, governing the speed of the game.

Both parameters relate to game difficulty, hence a significant impact on game engagement can be expected. As a matter of fact, providing a well-calibrated challenge is key to an engaging game experience. As phrased by Przybylski et al.~\citep{Przybylski2010}, the ``\textit{mastery of controls plays an important role in game motivation, largely as a necessary, but not sufficient, condition for achieving psychologically need-satisfying play.}''

\subsection{Summary}

The co-design activities presented in this section led to the design of a rehabilitation system, based on a new gesture-based input device and the game of Pac-Man. We have conducted preliminary informal testing, including with participants having sustained a SCI. 

In the remaining of the paper, we consider the problem of building a computational model of user performance while using the presented system. As introduced earlier, our motivation is to provide a more systematic method than heuristics in order to help adjusting the interaction parameters to define the game difficulty. Our approach relies on the estimation of the distance between a current performance and a baseline performance (reference). We split the problem in two: 
\begin{enumerate}
    \item Understanding the impact of the identified parameters on non-disabled participants' performance while using the game. The fact to perform the study on non-disabled participants is motivated by the need to set a baseline performance and how parameters affect this baseline performance (as further developed in Section~\ref{sct:experimental_study})
    \item Proposing a behavioural model able to predict participants' performance with respect to a baseline (Section~\ref{sct:modelling_user_behaviour}). 
\end{enumerate}

%% file: sections/experiment.tex

\section{Experimental Study}\label{sct:experimental_study}

In this section we investigate the impact of the identified parameters on non-disabled participants' performance while using the game. In addition, we collect usage data from the same population in order to set a baseline performance that will be further used in computational modelling (Section~\ref{sct:modelling_user_behaviour}).

\subsection{Apparatus}
The system described above was used for the experiment. The game was instrumented to expose internal state variables: the commands received; the score; and the number of the steps taken (in multiples of 10). These were logged along with their frame number and wall-clock timestamp.
The time rate control was implemented by fixing the duration of a game frame, equivalent to manipulating the frame rate.
The movements made by players, tracked by the position of the motion capture markers, were also recorded alongside game state variables.

\subsection{Task}
We used the first level of Pac-Man as the task. In this game, the Pac-Man needs to consume all the pellets organised in a maze-like 2-dimensional rectangular map. The maze is populated with four ghosts (Blinky, Pinky, Inky and Clyde) which generally chase Pac-Man and kill him on contact. Four power-ups, placed at the four corners of the maze, grant invincibility to Pac-Man for a short period of time. The task was completed when all pellets were consumed or when all three lives had been used.

\subsection{Design}
We use a  factorial design to measure potential effects and interactions of the independent variables gamepad control position (\SPREAD{}) and time rate (\TRATE{}) on score (\SCORE{}). \SPREAD{} was evaluated with two levels ($10cm$ and $40cm$)  and \TRATE{} was evaluated with three levels of slowdown ($T_r/3$, $2T_r/3$ and $T_r$). We opted for a within-subject design, as individual participants could be expected to have high variance. To counterbalance learning effects we used a balanced Latin square design. With independent variables \SPREAD{} and \TRATE{} evaluated with respectively two and three levels, we have six testing conditions.

To establish a reference gameplay experience, a pre-test and a post-test were included where participants used ordinary keyboard (\KEYBOARD{}) input to  control the Pac-Man game. This provides the reference data we later use to build a statistical model of normative gameplay. The pre-test and post-test involved playing two rounds of Pac-Man.

Each block was designed with three repetitions of the task, provided the third repetition (e.g. game) was started before the 8 minute mark in that block. When it was not the case, only one or two games would be played to completion, this to ensure the experiment would last one hour maximum. The design of the experiment comported thus: 12 participants $\times$ (2 \SPREAD{} $\times$ 3 \TRATE{} $\times$ 1 {\small{BLOCK}} $\times$ [1-3] Repetitions $+$ 2 \PRETEST{} $+$ 2 \POSTTEST{}) = [120-264] total trials.

\subsection{Procedure}
Participants were welcomed and asked to provide written consent after reading an information sheet. Upon acceptance, they received \pounds5 compensation. Participants were asked for demographic information and about their previous experience with \textit{Pac-Man}. They were given a short explanation about the game mechanics.
They were seated in front of a table, introduced to the tracking system, and equipped with a pair of headphones. An optical marker was attached with a stretchable band adaptable to different users' hand physical features. They were asked to acquire each of the gamepad controls once with the optical tracking system. The origin of the gamepad was chosen as a full forearm extension from the table edge adjacent to their trunk and, laterally, equidistant from both table edges. It was indicated on the tabletop with a protruding marker that taped in place for each participants. The participants were encouraged to produce their best possible score. To produce incentive for performance, we added an extra prize of \pounds15 awarded to the participant with the highest score in any of the testing conditions. At the end of each testing conditions (a combination of $r$ and $T_r$), participants were offered a break. Participants were responsible for advancing through the experiment by starting a new game. They were informed when they were changing testing condition but were not told about the current value of \SPREAD{} or \TRATE{}. At the end of the experiment, participants were invited to provide qualitative feedback during a short debriefing discussion.

\subsection{Participants}
12 participants served in the experiment: four females and eight males, one of whom was left-handed, with a mean age of $29\pm7$. The study was approved in advance by the University of X ethics committee. No participants presented with any relevant mental or physical disabilities. 

\subsection{Data Processing}
For each testing conditions, we recorded the logs from the game of \textit{Pac-Man}. From these, end-task \SCORE{} was computed as the highest game score at trial completion. Pac-man scores are integers which increase as various on-screen items (pellets, powerpills, monsters) are consumed. Typical scores for one level of Pac-Man are in the range 3000-8000 points, with a theoretical maximum of 14800 points.

To produce a unit-less measure of score comparison between \TRACKER{} and \KEYBOARD{} conditions, the normalised value \NSCORE{} was computed as the ratio between a raw score and the mean score from the keyboard reference games:
\begin{equation}
NSCORE = \frac{SCORE}{mean(SCORE_{keyboard})}
\end{equation}
In other words, a mean value of 1 for NSCORE in the \TRACKER{} condition means a level of performance equivalent to that produced with a keyboard.

\subsection{Results}
\subsubsection{Learning Effects and Score Distribution}
All participants reported having played Pac-Man, even if some reported that they could not remember when. We looked for learning effect between the reference \PRETEST{} and \POSTTEST{} conditions on \KEYBOARD{}. An ANOVA on \SCORE{} showed no significant difference between pre-test and post-test games. The mean score for all \KEYBOARD{} games was 4390 (s.d. 950), see Table~\ref{tbl:pre_post_test}.

\begin{table}[!ht]
\centering
\begin{tabular}{l | r r : r r | r}
     & \multicolumn{2}{c}{PRE} & \multicolumn{2}{:c|}{POST} &\\ \hline
game \# & 1 & 2 & 3 & 4 & all \\ \hline
mean & 3697 & 3992 & 4795 & 5077 & 4390\\
std  &  978 & 1438 & 1529 & 1705 & 950
\end{tabular}
\caption{Mean value and standard deviation for \SCORE{} over \PRETEST{} and \POSTTEST{} levels and averaged over games (all) in the last column.}
\label{tbl:pre_post_test}
\end{table}

\begin{figure}[!ht]
\centering
\includegraphics[width=\figratio\linewidth]{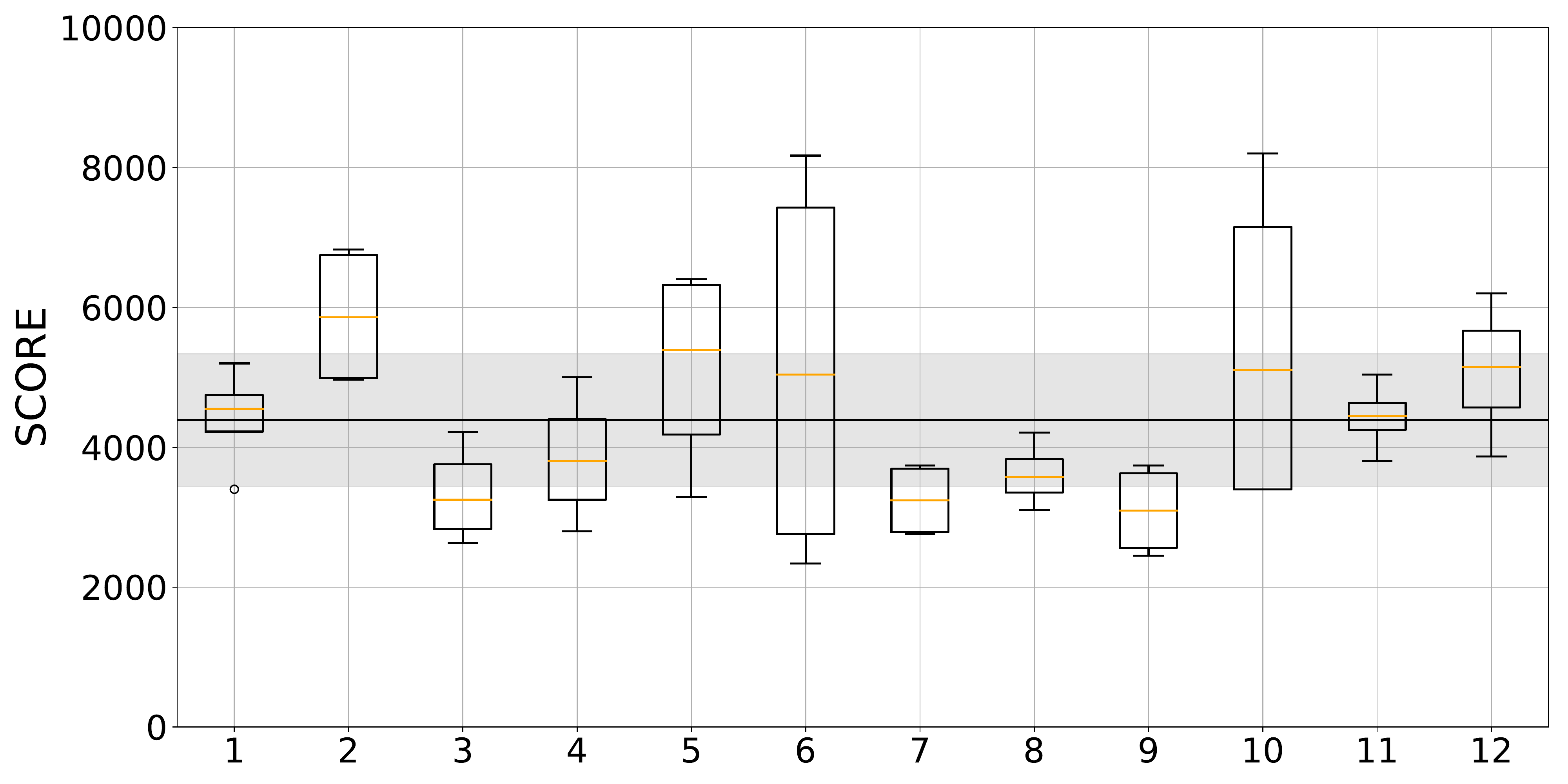}
\caption{Distributions of \SCORE{} across the 12 participants in the \KEYBOARD{} reference games. The mean and standard deviation across participants are plotted with a black line and a black shade, respectively.}
\label{fig:mean_scores}
\end{figure}

The distribution of \SCORE{} across participants was diverse (Figure \ref{fig:mean_scores}). Some participants, such as 1, 8 or 11, produced scores with a very low variance while others, such as 6 and 10, had a much bigger range of scores. 

\subsubsection{Effects on NSCORE}
\begin{figure}[!ht]
\centering
\includegraphics[width=\figratio\linewidth]{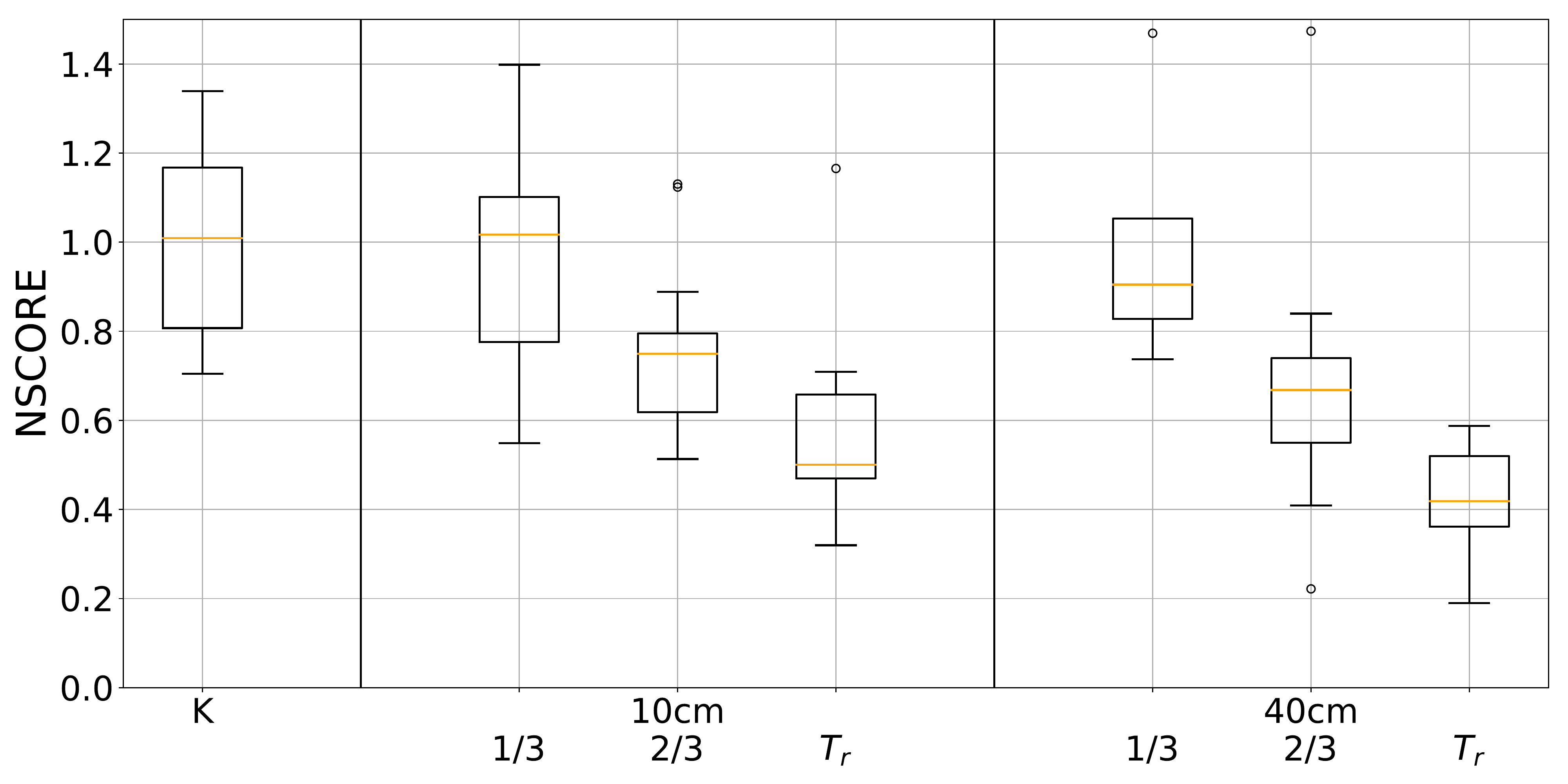}
\caption{Effects of \SPREAD{} and \TRATE{} on \NSCORE{} for \KEYBOARD{} (K) and \TRACKER{}. The graph includes three parts, with \KEYBOARD{} (K) on the leftmost subplot, and \TRACKER{} with the six level combinations grouped by value of SPREAD on the middle and rightmost subplots. Values for \SPREAD{} and \TRATE{} are reproduced on first row and second row, respectively.}
\label{fig:score_distribution_nscore_effect}
\end{figure}

The results for \NSCORE{} as function of \TRATE{} and \SPREAD{} are plotted in \reffigure{fig:score_distribution_nscore_effect}. A repeated measure two-way ANOVA on \NSCORE{} with \SPREAD{} and \TRATE{} as factors showed a significant main effect of \TRATE{} ($F_{1.47,16.15}=22.11$, $ges=0.41$, $p<0.0001$), no significant main effect of \SPREAD{} and no interaction \TRATE{} $\times$ \SPREAD{} on \NSCORE{}. \NSCORE{} was negatively correlated with \TRATE{}, decreasing from, on average, $96\%$ of the reference keyboard \SCORE{} at $T_r/3$ to $57\%$ at $T_r$.
We ran pairwise comparisons (adjusted with Holm-Bonferroni) on \SCORE{} between all levels with the inclusion of \KEYBOARD{}. It showed no differences of the \NSCORE{} means between \KEYBOARD{} and the condition where \TRATE{}  equals to $T_r/3$.

Supporting the lack of significant effect of \SPREAD{} on \NSCORE{}, we found no differences between any \SPREAD{} values for the same \TRATE{}. We also found no differences between levels ($T_r,10cm$), ($2T_r/3,10cm$) and ($2T_r/3,40cm$). In other words, participants performed as well using tracker as with a keyboard, measured by \NSCORE{}, provided \TRATE{} was reduced accordingly. The range of \TRATE{} values that produced a score  equivalent to a keyboard score is however unknown, but most likely centred around a value of \TRATE{} close to $T_r/3$. 
This result validates \TRATE{} as a suitable parameter to balance task difficulty as induced by SPREAD with game performance as measured by \NSCORE{}. For each value of SPREAD there exists a value for \TRATE{} such that \NSCORE{} under this condition is statistically indistinguishable from reference gameplay.

\subsection{Summary}

This experimental study explored the impact of input modality and design parameters $r$ and $T_r$ on game scores. We showed that for low speed the scores are not affected by the change of input device. As pointed out in the preliminary testing, the interaction with our system is more difficult than with the reference input device. This result shows that it is possible to counteract this effect, as measured by the game score, by reducing the time rate parameter $T_r$. In addition, we found that the spread does not affect the game score. 

With the objective of adapting the interaction such as ensuring a similar score than the one obtained using the baseline gameplay (reference input device and original speed), a metric must be defined to assess the performance between a given design and the reference one. Using the score as measured in this study is problematic. Since score is stochastic (people never play with the same exact performance twice), its reliable measurement requires a number of samples that increases with its variance. In addition, the value for score is sparse and delayed; available only once at the end of a game. Jointly, these factors limit the usability of score as a reliable, low-latency predictor of game performance due to its significant variation, high inter-subject variability and high latency.

In the next section we present a behavioural model and associated predictor for performance that is shown to be a good proxy for game score, to have lower variance than score and that can be computed with lower latency.

%% file: sections/model.tex

\section{Modelling User Behaviour}\label{sct:modelling_user_behaviour}

We propose to investigate probabilistic modelling of observed behavioural characteristics during gameplay. The goal is, first, to fit a probabilistic model onto baseline behavioural data. Second, we propose to use the trained model to assess the likelihood of observed behavioural data, when changing the input modality or design parameters, under the model.

\subsection{Behavioural Features}
Prior to modelling, we need to specify which behavioural features to consider. Some domain knowledge can be needed to chose these features. Here, our reference model includes the actions the players produces, measured by the interkey interval (\IKI{}), and the resulting effect in the game, measured by Pac-Man's turning time (\PTT{}). \IKI{} is computed as the number of frames between two consecutive issued commands. \PTT{} is computed as the number of frames in which Pac-Man stand motionless between two consecutive turns.  While \IKI{} is common in the literature related to typing~\citep{Dhakal2018}, the choice of \PTT{} has its roots from insights provided by game sessions with the prototype and backed by the experimental qualitative feedback: when poor control is afforded, producing rapid turns with Pac-Man becomes really hard. \IKI{} and \PTT{} features can be computed independently on the input modality considered (keyboard or tracker).

\subsection{Gameplay Reference Model}
The data needed to fit a reference model can be collected from baseline sessions. These baseline sessions have been recorded to that effect during the \PRETEST{} and \POSTTEST{} conditions on \KEYBOARD{} in the previous experiment.

We made the following assumptions. Both \IKI{} and \PTT{} were modelled as continuous random variables. Even though they take discretised positive values measured in frames, they represent a time measurement. We also assumed that \IKI{} and \PTT{} generated independent events at some constant average rate.

Under these assumptions, we propose to model the variable $X_{IKI}$, representing a time difference between independent events, by a Gamma distribution. The probability density function (pdf) is given by:
\begin{equation}
Gamma(x;k,\mu,\gamma) = \frac{(x-\mu)^{k-1} e^{-(x-\mu)/\gamma}}{\Gamma(k) \gamma^k}
\end{equation}
for $x \geq \mu$, $k > 0$, $\theta > 0$. $\Gamma(k)$ refers to the gamma function. We used maximum likelihood estimation (MLE) to estimate the distribution parameters. We fit the distribution on the \IKI{} features from the dataset of all trials from \PRETEST{} and \POSTTEST{} (keyboard condition),  across all participants. We found the following parameter values:  $k=2.19$, $\mu=-2.06$, $\gamma=17.11$.

\begin{figure}[!ht]
\centering
\includegraphics[width=\figratio\linewidth]{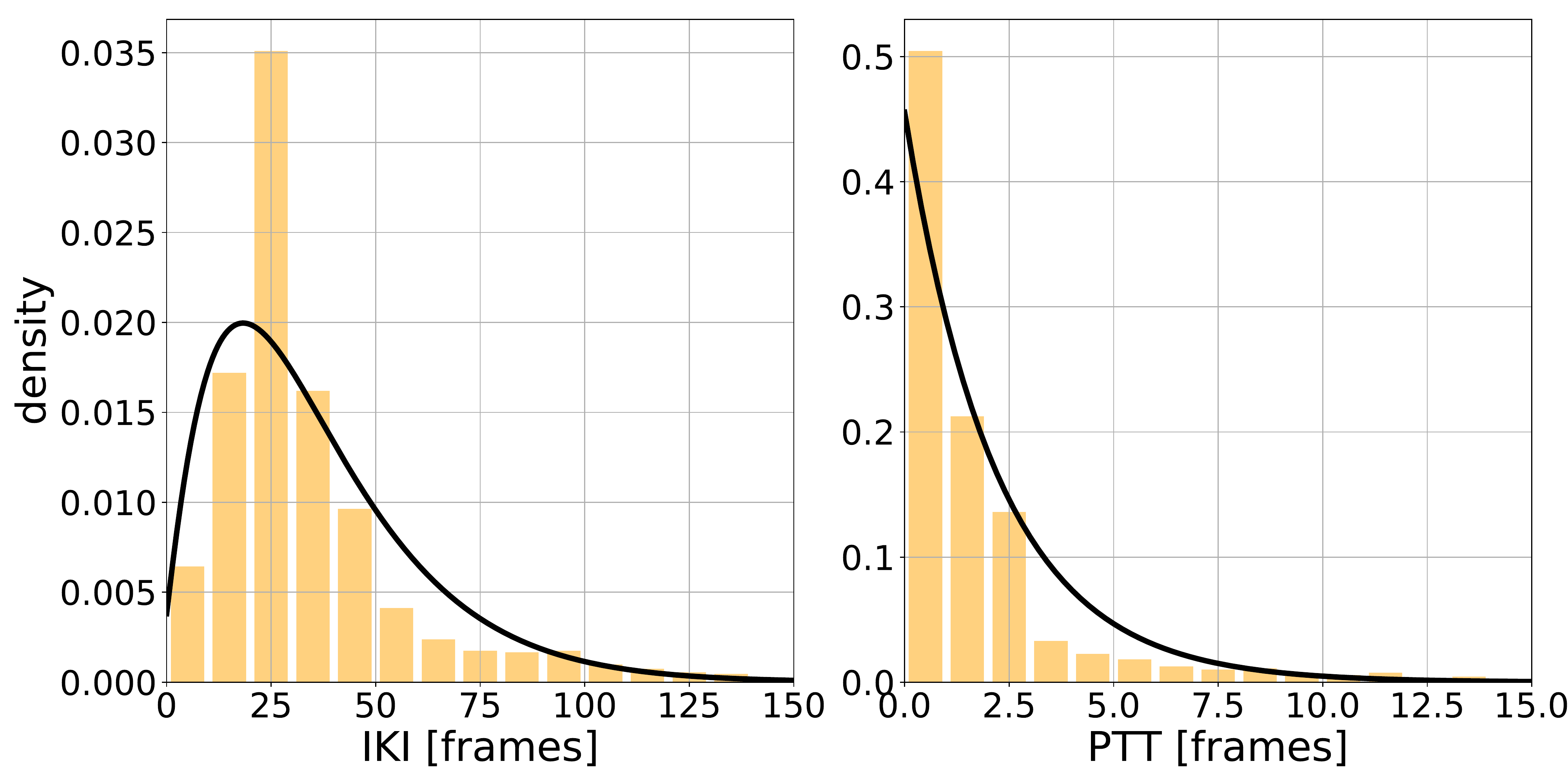}
\caption{Empirical (yellow histogram) and fitted (black pdf) distributions for variables PTT and IKI on the left and right plots, respectively.}
\label{fig:reference_distribution}
\end{figure}

For the variable $X_{PTT}$, the shape of the histogram exhibits a sharp peak around 0 with a rapid decay, which is similar to an exponential distribution (\reffigure{fig:reference_distribution}, right). Its probability density function (pdf) is given by:
\begin{equation}
Exp(x;\lambda) = \lambda \exp(-\lambda x),\ x \geq 0
\end{equation}
Similar to previously, we estimated $\lambda$ using MLE on the \PRETEST{} and \POSTTEST{} data, across all participants, and found that $\lambda=2.2$ provides the best fit.
Empirical distributions and the fitted models are shown in \reffigure{fig:reference_distribution}. Histograms (yellow bars) represent the empirical distributions, with their associated pdf shown in black.

Our reference model $\theta_{ref}: (k_{IKI}, \mu_{IKI}, \gamma_{IKI}, \lambda_{PTT})$ is fully defined by this quadruplets of parameters. Given the observed play session of a user, this model can be used to estimate the similarity between the observed user behaviour and the baseline keyboard behaviour.

The observations of $(X_{IKI}, X_{PTT})$ over $F$ frames  are gathered into $\mathbb{X}: (x^{IKI}_t, x^{PTT}_t)$.
We propose to use the likelihood $p(\mathbb{X}|\theta_{ref})$ of the observed data given our reference model $\theta_{ref}$ (keyboard-based reference behaviour) as a measure for inferring how likely a user is to be behaving as if interacting in the baseline condition.

Assuming independence between $X_{IKI}$ and $X_{PTT}$, we have:
\begin{eqnarray}
\begin{aligned}
p(\mathbb{X}|\theta_{ref}) & = p(X_{IKI}, X_{PTT}|\theta_{ref}) \\
&= p(X_{IKI}|\theta_{ref})p(X_{PTT}|\theta_{ref})
\end{aligned}
\end{eqnarray}
In practice we are taking the logarithm of this likelihood. We are interested in estimating the expected value of the log-likelihood, but we do not have access to the underlying $p(\mathbb{X})$ distribution. However, we can obtain samples from it and use the average value as an estimate:
\begin{eqnarray}\label{eqn:ll}
\begin{aligned}
\mathbb{E}[\log(p(\mathbb{X}|\theta_{ref}))]
\approx 1/F \sum_t^F \left[\log(p(x^{IKI}_t|\theta_{ref})) + \log(p(x^{PTT}_t|\theta_{ref}))\right]
\end{aligned}
\end{eqnarray}

\subsubsection{Normalisation}
The values for the log likelihood (\LL{}) were computed according to Equation \ref{eqn:ll}. In the same fashion as \SCORE{}, one value per game was obtained. We also computed a normalised value for \LL{}, marked as \NLL{} in the following, by taking the inverse of \LL{} and multiplying it with its average value obtained over the \PRETEST{} and \POSTTEST{} levels, \NLL{}$=mean($\LL{}$)/$\LL{}, in a similar fashion to how \NSCORE{} was computed.

\subsection{Comparing NLL to NSCORE}
\label{sct:nscore_nll}

\begin{figure}[!ht]
\centering
\includegraphics[width=\figratio\linewidth]{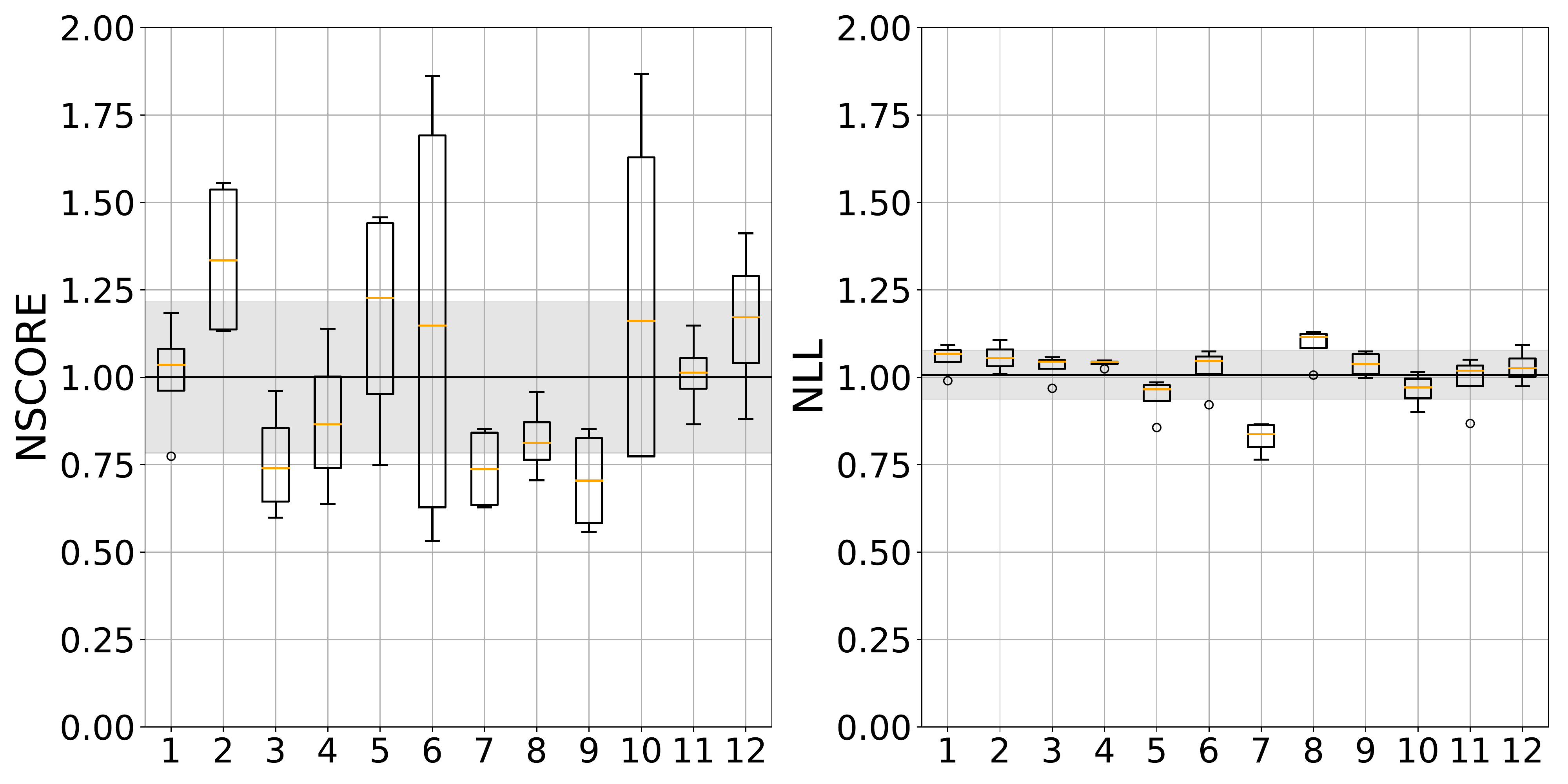}
\caption{Distributions of \NSCORE{} and \NLL{} for all participants over the \PRETEST{} and \POSTTEST{} conditions on the left and right, respectively. The mean and standard deviation of their distributions across participant are represented by a black horizontal line and a shaded horizontal area.}
\label{fig:nll_distribution}
\end{figure}

We inspected the likelihood of \IKI{} and \PTT{} observations per participant from \PRETEST{} and \POSTTEST{}.
\reffigure{fig:nll_distribution} reports the results.
We observed a clear difference in standard deviations, within and across participants, between \NSCORE{} and \NLL{} with a value of 0.22 and 0.06, respectively. For \NSCORE{}, five participants (2, 3, 5, 7 and 9) have an average \NSCORE{} which lies further than one standard deviation from the overall mean, while for \NLL{} only two participants (7 and 8) present the same deviation from the mean. This shows that \NLL{} is less subject to inter-user variability than \NSCORE{}.

\begin{figure}[!ht]
\centering
\includegraphics[width=\figratio\linewidth]{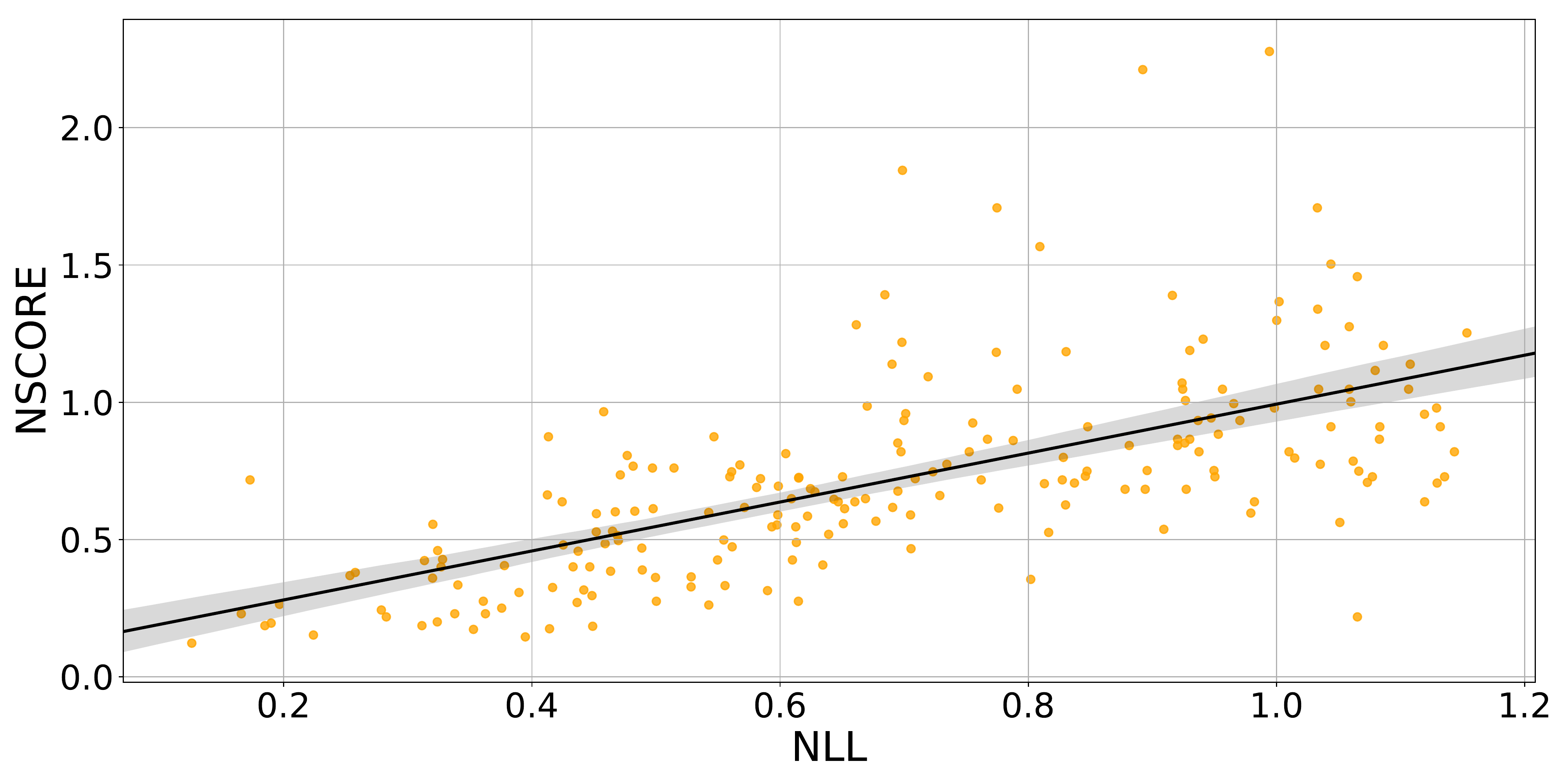}
\caption{Linear relationship between \NLL{} and \NSCORE{} with a Pearson correlation coefficient of $0.62$.}
\label{fig:correlation_evolution}
\end{figure}

Then we inspected the relationship between \NLL{} and \NSCORE{}. We computed the Pearson correlation through linear regression between \NLL{} and \NSCORE{}. The test revealed a significant correlation of $0.62$ ($p<0.001$, with $slope=0.83$, $intercept=0.14$, $stderr=0.07$), leading to $39\%$ of explained variance. A scatter plot (\reffigure{fig:correlation_evolution}) of their associated values for all measures except for the keyboard condition illustrates this relationship.

This correlation indicates that the statistical model based on the frequency of command inputs and their effect on game states is, to some extent, an indicator for user performance. The statistical model does not capture all the information necessary to predict score, such as the Pac-Man position in the maze or an understanding of the player's tactics. However, it is logical that the user ability to issue commands and control the avatar, modelled by \IKI{} and \PTT{}, is a factor influencing success in the game.

In order to inspect the contribution of \IKI{} and \PTT{} in the correlation between NLL and NSCORES, we computed the correlation of the likelihood using only one of the variables \IKI{} and \PTT{}. We found that they had different linear relationship with \NSCORE{}. A model with \IKI{} only showed a statistically significant correlation of $0.55$ ($p<0.001$, linear relationship with $slope=1.85$, $intercept=-0.94$, $stderr=0.19$) and a $R^2$ value of $0.30$. A model with \PTT{} only also showed a statistically significant correlation of $0.56$ ($p<0.001$, linear relationship with $slope=0.50$, $intercept=0.44$, $stderr=0.07$) and a $R^2$ value of $0.31$.

\subsection{Effect of design parameters on NLL}

Previous results showed that \NLL{} exhibits a linear relationship with \NSCORE{}. Here we inspected whether the design parameters (spread and time rate) impacts \NLL{}. To do so, we computed \IKI{} and \PTT{} features from the data collected using both the keyboard input modality and the movement-based input modality.

\begin{figure}[!ht]
\centering
\includegraphics[width=\figratio\linewidth]{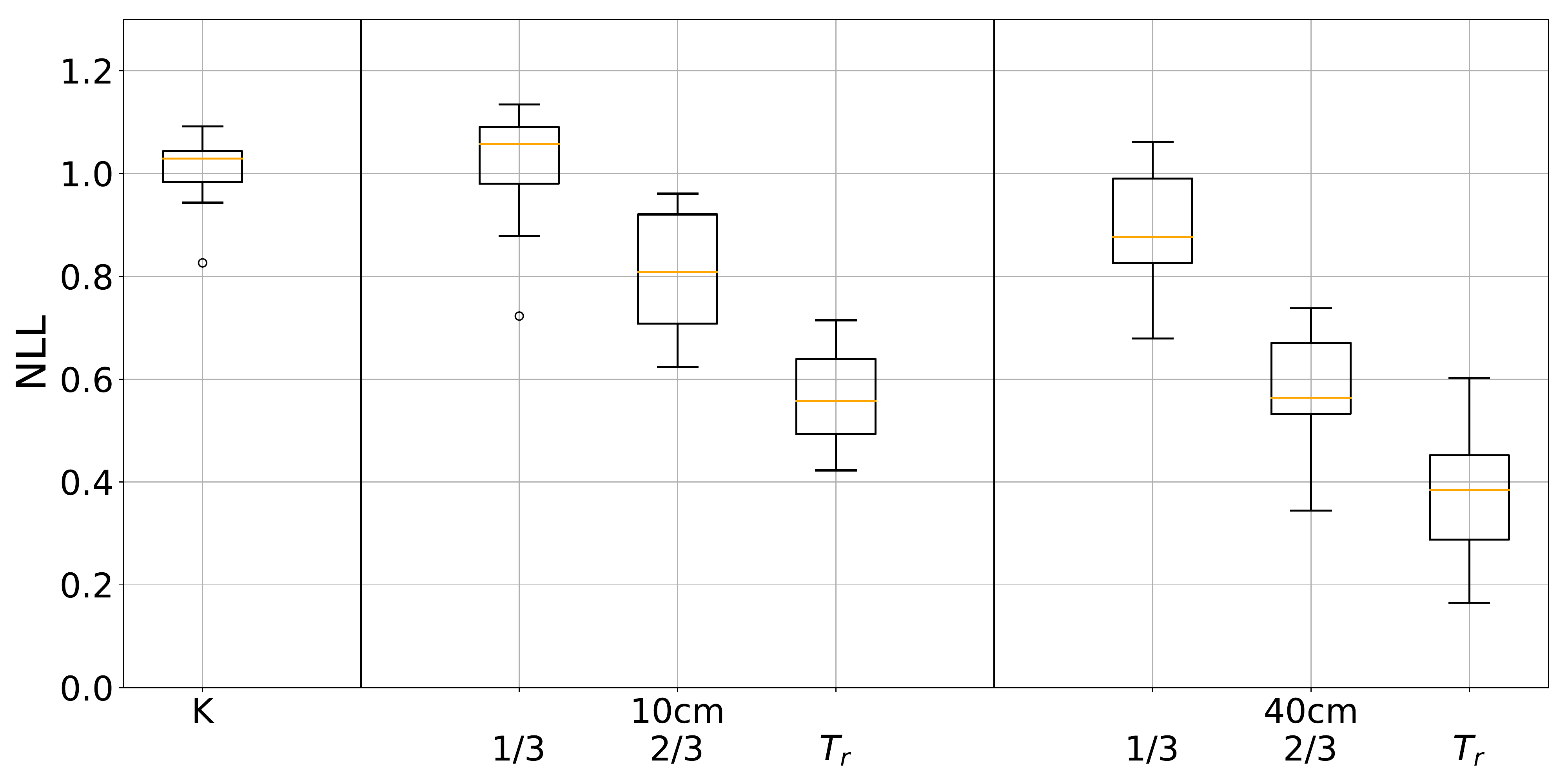}
\caption{Effects of \SPREAD{} and \TRATE{} on \NLL{} for \KEYBOARD{} (K) and \TRACKER{}. The graph includes three parts, with \KEYBOARD{} (K) on the leftmost subplot, and \TRACKER{} with the six level combinations grouped by value of \SPREAD{} on the middle and rightmost subplots. Values for \SPREAD{} and \TRATE{} are reproduced on first row and second row, respectively.}
\label{fig:nll_effect}
\end{figure}

\reffigure{fig:nll_effect} depicts the statistics computed on \NLL{} values for each condition. The figure reports the likelihoods computed using data from the movement-based input modality under the six testing conditions in the middle and rightmost columns.

We ran a repeated measure two-way ANOVA on \NLL{} with \SPREAD{} and \TRATE{} as factors. The analysis showed a significant main effect of \TRATE{} ($F_{1.96,21.58}=83.34$, $ges=0.76$, $p<0.0001$). The analysis also showed a significant main effect of \SPREAD{} ($F_{1,11}=72.50$, $ges=0.39$, $p<0.0001$). We observed a negative impact of \TRATE{} and \SPREAD{} on \NLL{} with lower values with increasing values of \TRATE{} and \SPREAD{}. We ran pairwise comparisons adjusted with Holm-Bonferroni on all levels with the inclusion of \KEYBOARD{}. We found that only level ($T_r/3$, $10cm$) was similar to \KEYBOARD{}. We also found some similarities between pairs ($T_r/3$, $10cm$) and ($T_r/3$, $40cm$) and pairs ($T_r/3$, $40cm$) and ($2T_r/3$, $10cm$) showing that similar effect on \NLL{} can be achieved with different parameters pairs. Finally, the analysis found a small interaction \TRATE{} $\times$ \SPREAD{} ($F_{1.58,19.39}=5.28$, $ges=0.04$, $p=0.02$). The reason seems to be the relatively high value of \NLL{} for level ($T_r$, $40cm$). However, the amount of data collected in this condition is comparatively smaller than other conditions: the poor user performance had the result of shortened game sessions. This effect is thus likely an artefact.

These results should be compared with the effect of \SPREAD{} and \TRATE{} on \NSCORE{} (\reffigure{fig:score_distribution_nscore_effect}). The goal of providing an experience similar to the keyboard reference translates to having parameters value ($r$ and $T_r$) producing \NSCORE{} and \NLL{} equal to unity. The statistical analysis showed that the parameters value ($10cm$, $T_r/3$) is a solution for our design, whether \NSCORE{} or \NLL{} is considered. In other words, two models of user behaviour, based on widely different measurements (the outcome of the interaction for \NSCORE{} and realisations of low-level variables for \NLL{}), did identify the same value for $T_r$ to produce an interaction similar to a keyboard. 
There are also some differences between both models. For example, the parameters value ($40cm$, $T_r/3$) is also a solution for the model based on \NSCORE{} but not for \NLL{}. The fact that \NLL{} presents a much lower variability than \SCORE{} could explain this difference.

\subsection{Sampling Period}
\begin{figure}[!ht]
\includegraphics[width=\figratio\linewidth]{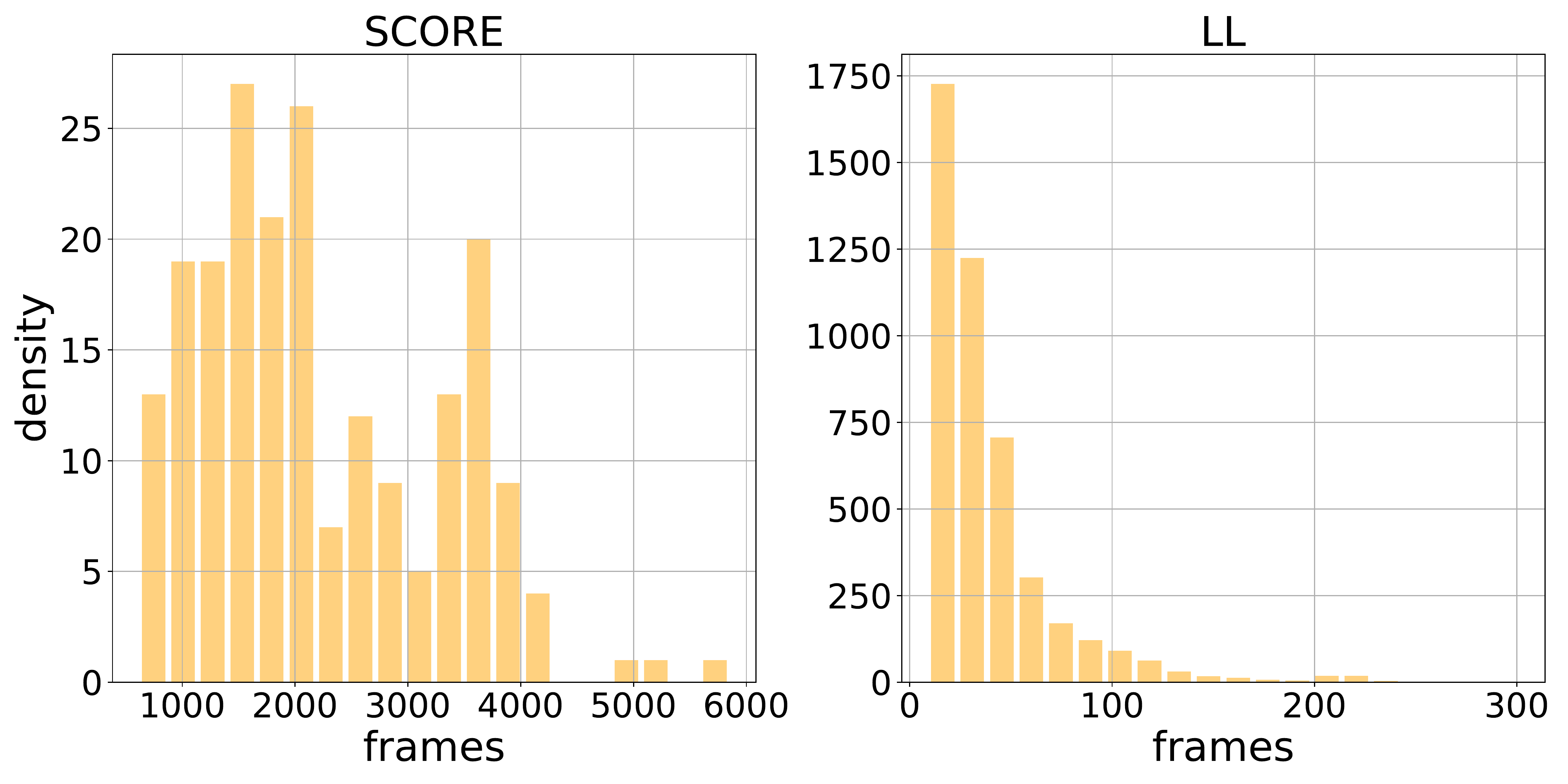}
\caption{Distribution of time taken to measure a sample from \SCORE{} and \LL{} on the left and right, respectively.}
\label{fig:score_ll_avail}
\end{figure}

Finally, one of the reason for designing such model was the availability of many more samples for the variables \IKI{} and \PTT{} than observations of \SCORE{} per unit of time. We measured from the \TRACKER{} condition the time elapsed between samples for \SCORE{} and \LL{} (\reffigure{fig:score_ll_avail}). Obtaining one sample for \SCORE{} took on average $2210$ (s.d.$1057$) frames, while one sample for \LL{} was available every $40$ (s.d.$31$) frames on average. The measure of user behaviour based on low-level variable of gameplay provides $55$ times more samples per frame than the counterpart model based on \SCORE{}, thus exhibiting a lower latency.

%% file: sections/discussion.tex
\section{Discussion}

This paper contributes to research in interactive systems for motor rehabilitation. 
We presented two main contributions. First, we presented a gesture-based interactive system, interfacing with a commercial game, specifically designed such as complying with requirements from the context. These requirements were provided by occupational therapists involved in a co-design process. One of these requirements was to be able to adapt the level of the game difficulty through interaction parameters, that could be set by the OTs or automatically adapted. Our second contribution is to identify these parameters and propose a computational model estimating, in realtime, the input control performance as compared to a baseline performance, which could then be used as a proxy for game performance or optimisation function to adapt the parameters automatically. In this section, we discuss the different findings.

\subsection{Design workshop and tools for modelling}

The co-design activities (section \ref{sct:motivation}) with occupational therapists were effective in exploring interaction possibilities and establishing requirements for a gamified interaction of the upper limb for patients with SCI.

The design of a new control modality was the principal outcome of the workshops with OTs. For motor rehabilitation, which targets specific motions from patients, the input control is the principal interface between the user's body and the game. This is quite different from other types of rehabilitation targeting for example cognitive impairments~\citep{Mandryk2013}. As a result, the game itself or the choice of the game did not emerge as a priority. Instead, the OTs mentioned the idea of reusing the proposed design with different games for satisfying specific patient preferences, a practice also described in the work by Hofmann et al.~\citep{Hofmann2019} as design reuse across clients.

Supported by rapid design iterations with a paper prototype, an adequate mapping from desired movement space to game control space~\citep{Walther-Franks2013a} was identified as well as a suitable parameterisation for adapting task difficulty~\citep{Lopes2011}. The physical separation between controls placed on a tabletop effectively increased the reaching motions, in accordance to the rehabilitation requirements. This was helped by the natural connection between reaching motions and targeting motions but remains an issue for other types of activity. For more complex mapping, such as the one involving pedalling on a recumbent bicycle proposed by Ketcheson et al.~\citep{Ketcheson2016}, an obvious solution to the mapping task is hard to obtain.
We identify here potential opportunities for linking serious game design research with research related to HCI, such as~\citep{Bachynskyi2015a, Oulasvirta2013}. The breadth of possibilities afforded by the design of a new input control, and the effects these have on the user's body are not yet fully understood. 

As for the game adaptation, we chose not to use additional graphical overlays~\citep{Ketcheson2016}. 
Instead, we relied on the flexibility of the control modality via the parameter spread and on the well-known game design pattern using time rate fluctuations. The OTs' definition of the targeted motions was not very restrictive, on the contrary to systems that target more precise motion executions, such as \textit{YouMove}~\citep{Anderson2013} or \textit{Physio\@Home}~\citep{Tang2015}, wherein the use of graphical feedback to convey how well the user is performing motions seems hardly avoidable. Put together, this affords a simple parameterisation, with the spread and time rate mainly controlling to rehabilitation goals and gameplay performance, respectively.

\subsection{Interaction parameters adaptation}

Game interaction with modified control input requires adaptation and usually results in (much) harder challenge for the player. Apart from controllers designed for performance, such as the one proposed by Kwak et al.~\citep{Kwak2009}, new input controls can not afford users the same proficiency.

This effect was measured during the experimental study (section \ref{sct:experimental_study}) by including a control condition and evaluation against a measure of in-game score.
We showed that the parameter $T_{rate}$ was successful in adapting the interaction difficulty: reducing $T_{rate}$ from its original value decreased the level of difficulty, and with a preset level fixed around $1/3$ the scores were indistinguishable. This is close to the $1/4$ ratio in information throughput measured by Card et al.~\citep{Card1991} between the arm and the fingers.
This also quantifies the significant change in users' control capabilities when input devices are swapped, which is thus an important effect to quantitatively measure when similar systems are proposed.

However, finding approximate preset values for the interaction parameters so that users could enjoy a performance similar to the keyboard reference is time consuming when score is used as a metric due to its sparse availability and high variability. In addition, any further modifications to the design of the control input would render previous presets unusable, requiring another round of measurements. Finally, since the data collected involved only unimpaired participants, it is unlikely that these preset values transfer to the final target group. As a result, we developed a low-latency metric of baseline gameplay that could replace the reliance on score for that matter.

\subsection{Modelling Gameplay}

We proposed modelling baseline gameplay (section \ref{sct:modelling_user_behaviour}) using game-specific features and user input, and using the normalised log-likelihood of observed gameplay under the reference model as a similarity measure. We showed that this measure is highly correlated with score, has lower variance and is available at lower latency.

We relied on the reasonable assumption that unmodified play is engaging. Our model of gameplay thus captures the behaviour of users, through their input frequency or \IKI{}, and the effect they have on the game, through the specifically designed variable \PTT{}. Even though \PTT{} is hand crafted feature, it is simply an indirect measure of the avatar speed.

Then, we showed that this model correlates with score. This is evidence that maintaining gameplay, defined by \IKI{} and \PTT{}, is important to ensure game performance. The correlation between both metrics is however not perfect. 
The totality of the behaviours related to game outcome are not taken into account by the approach we proposed. Only the number of actions per unit of time and the speed of the Pac-Man in the game were considered. More complex models, such the one proposed by Smith et al.~\citep{Smith2016} could be employed if the objective is to predict with accuracy the final score. Complex models are however hard to train. For instance, Smith et al.~\citep{Smith2016} reported training duration in the order of days. The point here is that simple models, yet able to generalise, are valuable. The model we proposed relied on four parameters only, and was shown less prone to inter-user variability than in-game score.

Our results showed that the spread parameter does not impact the score.
We can speculate that a higher spread may have impacted the score values, or that the spread considered would have impacted the performance of participants with impairments. However the spread does impact the log-likelihood values, suggesting that the proposed model may be more suitable to assess the input control performance than game score. The spread parameter was meant to steer motions towards those targeted by OTs for rehabilitation. Having a model that is capable of leveraging changes in spread is therefore important from a rehabilitation point of view. 

Finally, a model linking engagement and performance is not easily established. Access to user engagement requires having specific measurements which often rely on questionnaires. Using the flow framework~\citep{Csikszentmihalyi2014}, Limperos et al.~\citep{Limperos2011} compared the difference a Wii and a Playstation controller produced on the game experience, with the Wii controller being qualified as ``natural'' and requiring more involvement from the user body. They showed that the performance as defined as the end-game score was not alone explaining the enjoyment, and that the sense of control that users experienced was a rather salient indicator. Understanding what part of the gameplay makes a game engaging is an interesting research question. There is a need for models that are interpretable and that rely on quantifiable gameplay measures to further the understanding of user engagement.

\subsection{Limitations and Future work}

We acknowledge several limitations to this work. 
We made several assumptions with regards to the model. One was that samples were independent of each others, and \IKI{} and \PTT{} were also independent. These simplification assumptions are useful to build a simple model: \IKI{}/\PTT{} independence allows to model them separately, otherwise a joint distribution should have been learned which would have required a more complex learning procedure. The assumed independence of consecutive samples drawn from \IKI{} and \PTT{} is however useful to use simple probabilistic distributions with rather good generalisation capabilities. It is justified here with regards to the variables themselves: they describe short term processes that may not vary much with time.

The simplicity of the game we used is both an advantage and a disadvantage. For games using a similar control scheme, this approach is likely to be transferable. For game with more complex control commands, more work should be done on movement representation and probabilistic modelling. The distributions that were modelled in this work were simple (i.e. unimodal) and stationary which might not be always be the case.

The baseline model we trained is meant to be representative of a general population of players. This is one of the reason for its simplicity and the focus on command frequency: the simpler the model, the lesser the risk of overfitting. With a total population of 12 participants however, it is likely that certain category of players would be incorrectly modelled. In that case, it could be possible that players performing well with our system while behaving out of the norm would be qualified as unlikely by our model.

Finally, as future work we plan to test the model we developed to optimise online the interaction parameters. This extension to our computational approach is qualified as computational design~\citep{Oulasvirta} and is a clear path forward for our research.

\section{Conclusion}
We have shown that a computational approach to design can be used as a mean to systematically control interaction performance in games where input device characteristics are varied. We argued that performance is an important side of the interaction to take into account when creating new control devices. We proposed a probabilistic model of low-level user behaviour that affords a measure of similarity against baseline gameplay, that could be used to online system adaptation. Through a complete methodology from field work to experimental and computational studies, this work brings a pragmatic account for designing interactive systems while maintaining user performance in a rehabilitation context. 

\section{Acknowledgements}
We would like to warmly thank our contacts from the Spinal Unit of the QEU hospital (\url{https://www.spinalunit.scot.nhs.uk/}), in particular Michelle Rankin and Mariel Purcell, for the time they spent sharing their valuable expertise, helping with the design and testing of the prototype and their overall support for this research, which was funded through the MOREGRASP project (\url{http://www.moregrasp.eu/}).